\documentclass[aps,prd,reprint,superscriptaddress,footinbib,twocolumn]{revtex4-1}

\usepackage{dsfont}
\usepackage{amsmath}
\usepackage{mathrsfs}
\usepackage{enumerate}
\usepackage{graphicx}
\usepackage{multirow}
\usepackage[latin1]{inputenc}
\usepackage{ulem} 
\usepackage{units} 
\usepackage{version} 
\usepackage{color}
\usepackage{xcolor}  
\usepackage[colorlinks=true,citecolor=black,urlcolor=black,linkcolor=black]{hyperref} 
\usepackage{array}

\newcommand{\m}{\ \text{m}}
\newcommand{\s}{\ \text{s}}

\newcommand{\Hz}{\ \text{Hz}}

\newcommand{\km}{\ \text{km}}

\newcommand{\sqrtHz}{/\sqrt{\text{Hz}}}
\newcommand\stxt[1]{_{\text{#1}}} 

\newcommand{\beq}{\begin{equation}}
\newcommand{\eeq}{\end{equation}}
\newcommand{\NNtype}{(u,a)}

\begin{document}



\title{Low Frequency Gravitational Wave Detection With Ground Based Atom Interferometer Arrays}

\affiliation{ARTEMIS,  Universit\'e C\^ote d'Azur, CNRS and Observatoire de la C\^ote d'Azur, F-06304 Nice, France}
\affiliation{LNE-SYRTE, Observatoire de Paris, PSL Research University, CNRS, Sorbonne Universit\'es, \\ UPMC Univ. Paris 06, LNE, 61 avenue de l'Observatoire, 75014 Paris, France}
\affiliation{LP2N, Institut d'Optique d'Aquitaine, CNRS, rue Francois Mitterrand, 33400 Talence Cedex, France}

\author{W. Chaibi$^{1}$}
\email{chaibi@oca.eu}
\author{R. Geiger$^{2}$}
\email{remi.geiger@obspm.fr}
\author{B. Canuel$^{3}$}
\author{A. Bertoldi$^{3}$}
\author{A. Landragin$^{2}$}
\author{P. Bouyer$^{3}$}


\date{\today}

\begin{abstract}
\label{abstract}

We propose a new detection strategy for gravitational waves (GWs) below few Hertz based on a correlated array of atom interferometers (AIs).
Our proposal allows to reduce the Newtonian Noise (NN) which limits all ground based GW detectors below few Hertz, including previous atom interferometry-based concepts.
Using an array of long baseline AI gradiometers yields several estimations of the NN, whose effect can thus be  reduced via statistical averaging.
Considering the km baseline of current optical detectors, a NN rejection of factor 2 could  be achieved, and tested with existing AI array geometries.
Exploiting the correlation properties of the gravity acceleration noise, we  show that  a 10-fold or more NN rejection is possible with a dedicated  configuration.
Considering a conservative NN model and the current developments in cold atom technology, we show that strain sensitivities  below $1\times 10^{-19}\sqrtHz$ in the $ 0.3-3 \Hz$ frequency band can be within reach, with a peak sensitivity of $3\times 10^{-23} \sqrtHz$ at $2 \Hz$.
Our proposed configuration could extend the observation window of current detectors by a decade and fill the gap between
ground-based and space-based instruments.

\end{abstract}

\maketitle
\section{Introduction}
\label{par:intro}

Gravitational Wave (GW) detection remains today one of the challenges in fundamental physics and astrophysics. State-of-the-art GW detectors consisting in giant Fabry-Perot Michelson interferometers \cite{CellaGiazotto2011, KAGRA2012,AdV2014,GEOHF2014,Collaboration2015}  now reach a sensitivity 
that justifies the expectations for a direct  detection of GWs in the next few years \cite{Abadie2010}. Nevertheless, low frequency GW sources will  remain hidden for ground based detectors for which the observation bandwidth will be limited
to frequencies above few Hz \cite{Punturo2010}. 
Still, reaching sub-Hz sensitivities could provide a decisive asset towards GW astronomy as the sources in this band produce more powerful and durable signals \cite{Harms2013}.
In this purpose, hybrid detectors based on two distant Atom Interferometers (AIs) interrogated by a laser propagating over a long baseline have been  proposed (see, e.g. \cite{Dimopoulos2008PRD}). Using as test masses free falling atoms  instead of suspended mirrors could resolve most of the technical limitations presented by optical GW detectors at low frequency, such as residual seismic noise or thermal noise of suspension systems.

Like all ground based detectors, current atom interferometry proposals will nevertheless suffer from the so-called Newtonian Noise (NN) \cite{Saulson1984}. NN consists in fluctuations of the terrestrial gravity field which creates a tidal effect on separated test masses and is indiscernible from the effect of a GW \cite{Saulson1984,Vetrano2013}. NN is therefore considered as a fundamental limit for any ground based GW detectors at frequencies below a few Hz. 
Various methods   have been considered to circumvent this problem \cite{eLisa2012,Driggers2012,HarmsHild2014,HarmsPaik2015}.
In this communication we propose a new concept which uses an array of AIs   configured to  reject the NN. 

\begin{figure}[!h]
  \centering
 \includegraphics[width=\linewidth]{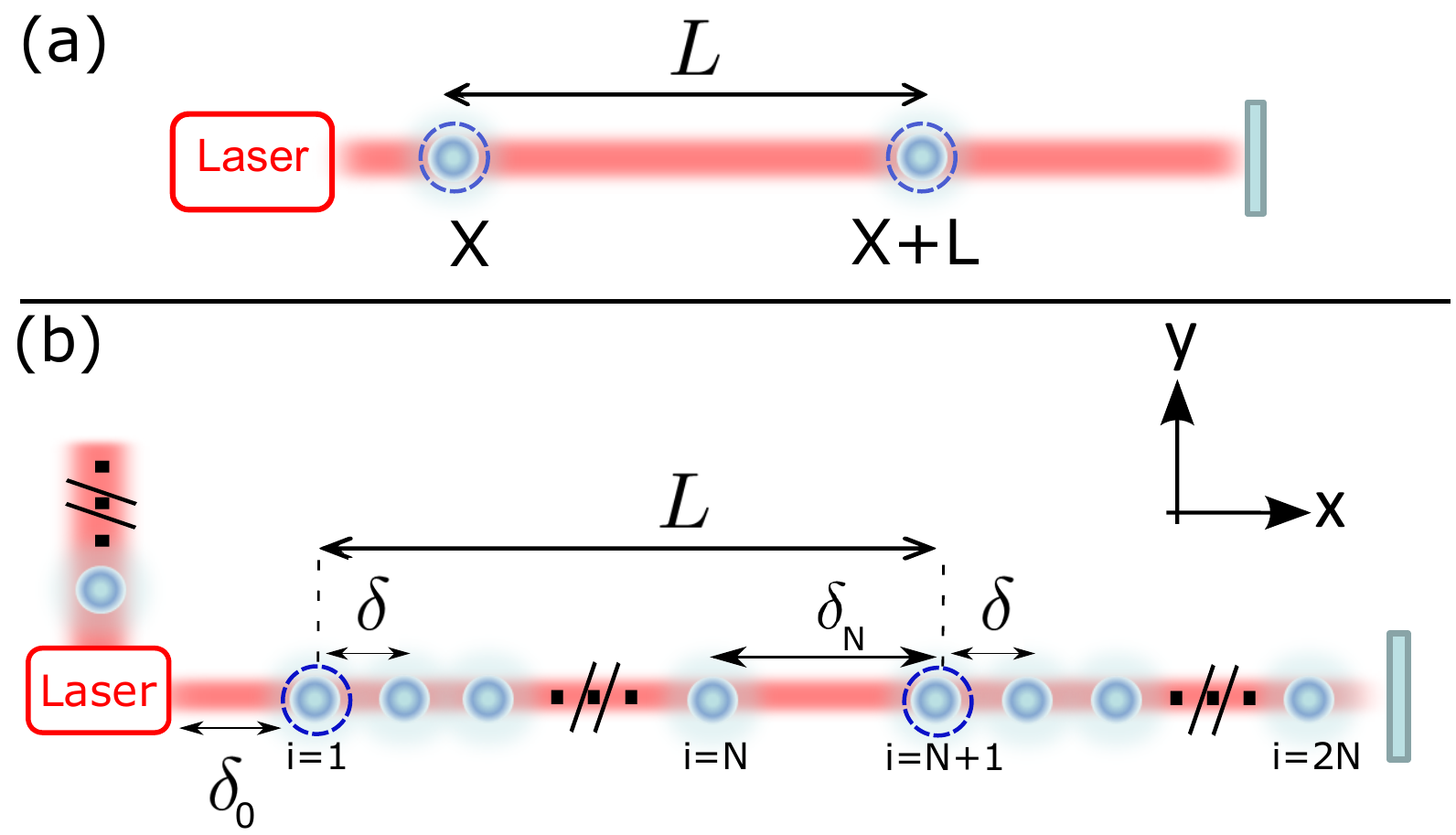}
  \caption{
(a) A single gradiometer using two AIs at positions $X$ and $X+L$, interrogated by a common laser beam. 
(b)  An array of $N$ AI gradiometers used for sampling the spatial variations of the NN. The separation  between the gradiometers is $\delta$.  The array allows repeating $N$ times the  experiment sketched in (a) and averaging the NN. 
The use of two orthogonal arms injected by a common laser enables to reject laser frequency noise (the second arm in the y direction is only partially represented here for clarity). 
  }
 \label{fig:schematic_detector}
  \end{figure}

Unlike previous single  strainmeter/gradiometer proposals where the NN and the GW signal are indiscernible, the array of AIs allows to extract the GW signal by averaging over several realizations of the  NN. 
The NN rejection can 
be further enhanced by exploiting the correlation spatial behaviour of the gravity acceleration.
With the $3-4 \ \text{km}$ baseline of current best optical detectors, our method reaches a NN rejection of about 2; this factor is comparable to what obtained other passive methods (e.g. \cite{HarmsHild2014}). 
The principle of such rejection can be tested in current AI array projects
\cite{MIGA2014iDust}.
We focus in this communication on a $16 \ \text{km}$ baseline detector that takes full advantage of our method
and enables strong NN rejection by more than a decade.
This could complement 
the current optical interferometers development program by opening the $\sim 0.3-3\Hz$ observation window.

\section{Principle}
\label{principle}
  
A single AI gradiometer (Fig.~\ref{fig:schematic_detector}a) consists of two AIs separated by a baseline $L$ 
and interrogated by a common laser beam of frequency $\nu$ close to the atomic transition frequency (see, e.g. \cite{Dimopoulos2008PRD}).
We consider a 3 light-pulse AI (with $T$  the time between the successive pulses) 
using Bragg diffraction of atoms
from a standing wave produced by retro-reflecting the 
interrogation laser. 
The output phase of 
each AI originates from the local  phase difference $\Delta\varphi$
between the two counter-propagating beams at the time of the pulse and the position of the atom \cite{Borde2004GRG}. 
The retro-reflection configuration gives immunity to laser phase noise induced by position noise of the input laser system.
 We consider Large Momentum Transfer (LMT) 
 diffraction \cite{Mueller2008}, where the atom absorbs $n$ photons from one beam and  emits $n$ photons in the 
 counter-propagating beam. 
 When the interrogation laser is pulsed, the atom undergoes diffraction with a momentum change of
$n\times 2\hbar k$ along the laser propagation direction ($k=2\pi\nu/c$ 
is the laser wave-vector).   
A phase $n\Delta\varphi\left(X,t\right)$  is imprinted  on 
the diffracted component.

After the 3 pulses, the output phase of the AI reads 
\begin{eqnarray}
\label{phase_AI}
& &\Delta\phi_x\left(X,t\right) = \epsilon\left(X,t\right)+2 n k\times\\
\nonumber & &\left[\left(\frac{\Delta\nu(t)}{\nu}+\frac{h(t)}{2}\right)\left(L-X\right)+\Delta x_2(t)-\Delta x\left(X,t\right)\right]\otimes s(t)
\end{eqnarray}
where $\epsilon\left(X,t\right)$ represents the detection noise (e.g. atom shot noise) on the output phase of an AI using atoms placed at position $X$. $s\left(t\right)$
is the sensitivity function of the 3 pulse AI 
\cite{Cheinet2008IEEE}
and  relates the AI output phase to the second temporal derivative of the 
local laser phase difference $\Delta\varphi$. 
 $\Delta x_2(t)$ is the position noise of the retroreflecting mirror and $\Delta x(X,t)$ 
represents the motion of the  atoms along the laser beam direction due to the fluctuations of the local gravitational acceleration.

Taking the differential phase $\psi(X,t)=\Delta\phi_x\left(X,t\right)-\Delta\phi_x\left(X+L,t\right)$ 
between two AIs separated by the distance $L$ and neglecting  laser frequency noise yields:
\begin{eqnarray}
\label{eq:gradient}
\nonumber \psi(X,t) &=& 2nk\left[\frac{L\ddot{h}(t)}{2}+a_x\left(X+L,t\right)-a_x\left(X,t\right)\right]\otimes s_\alpha(t)\\
&+&\epsilon\left(X,t\right)-\epsilon\left(X+L,t\right),
\end{eqnarray}
where $s_\alpha(t)$ is the AI sensitivity function to acceleration, given by $\ddot{s}_\alpha(t)=s(t)$.
Importantly,  position noise $\Delta x_2(t)$ of the retro-reflecting mirror has been rejected in this gradiometer configuration. Eq.~\eqref{eq:gradient} shows that  fluctuations of the local gravity field result in 
an acceleration signal $a_x(X,t)=\Delta\ddot{x}(X,t)$ whose gradient will have the same signature as that of the GW (see Ref.~\cite{Borde2004GRG} for a more rigorous calculation).

The differential phase of Eq.~\eqref{eq:gradient} can be written as $\psi(\tilde{\eta})=H(t)+\tilde{\eta}(t)$ where $H(t)$ is the GW signal and $\tilde{\eta}(t)$ the noise (detection noise and NN) at position $X$. 
Our idea is to extract $H(t)$ using a Monte-Carlo method:  the GW signal is obtained by averaging over several samples of the noise $\tilde{\eta}(t)$, 
which formally reads $H = \int \Psi(\tilde{\eta}) \text{d}\tilde{\eta}$.
To this aim, we consider $N$ realizations $\{\psi(X_i,t)\equiv\psi_i(t)\}_{i=1..N}$ of the single gradiometer 
and compute the average signal
\begin{equation}
\label{average}
H_N\left(t\right) =\frac{1}{N}\sum^{N}_{i=1} \psi_i(t),
\end{equation}
which represents a non biased approximation to the GW signal of interest, i.e. $Lh\left(t\right)/2$. 
Assuming that the $N$  realizations are independent, the residual noise on the GW measurement is reduced by $\sqrt{N}$,
\begin{equation}
\sigma_{H_N} = \frac{\sqrt{2}\sigma_\eta}{\sqrt{N}},
\label{eq:MCnoisered}
\end{equation}
with $\sigma_\eta=\sqrt{\sigma_a^2 + \sigma_\epsilon^2}$  the standard deviation (s.d.) resulting from the  NN and detection noise which we considered as independent variables of s.d. $\sigma_a$ and $\sigma_\epsilon$, respectively.
We assumed uncorrelated noise between the 2 AIs of a single gradiometer, yielding the $\sqrt{2}$. This is always valid for the detection noise and  applies for the NN when the gradiometer baseline $L$  is much larger than the NN correlation length.
Since the GW signal increases with $L$, a very long gradiometer baseline will be considered in the following, which validates the assumption of uncorrelated NN bewteen the two AIs.
As the $N$ gradiometer measurements have been assumed independent, the AI array brings to a $\sqrt{N}$ rejection factor for the  NN (and for the detection noise).

We study an  implementation of this Monte Carlo sampling method in which $N$ different gradiometer measurements are simultaneously realized in parallel thanks to an array of spatially distributed AIs. The proposed configuration is chosen to 
enhance the NN reduction via
variance reduction 
\cite{Caflisch1998}. For that, we optimize the AI array distribution, i.e. the signal spatial sampling, in order to benefit from the spatial behavior of NN correlations. 
We show that, in a given frequency band, a significant additional rejection factor can be gained with respect to the standard $\sqrt{N}$ of Eq.~\eqref{eq:MCnoisered}.

\section{Implementation and sensitivity of the detector}
\label{implementation}
The implementation is sketched in Fig.~\ref{fig:schematic_detector}(b).
We consider a symmetric configuration consisting of 2 orthogonal arms of same length and interrogated by the same laser. For a GW with (+) polarization, laser frequency noise is therefore rejected (see Appendix \ref{ap:seismic_iso} for more details).
Each arm of total length $L_a$ consists in a series of gradiometers of baseline $L=X_{N+i}-X_i$  which are  separated by the distance $\delta$. The geometrical parameter $\delta_N$ reflects that the baseline $L$ and the  separation $\delta$ between the gradiometers are independant. 
For $1\leq i\leq N$ we define
\begin{eqnarray}
\label{differential}
\nonumber \psi_i\left(t\right)&=&\left[\Delta\phi_x\left(X_i,0,t\right)-\Delta\phi_x\left(X_{N+i},0,t\right)\right]\\
&-&\left[\Delta\phi_y\left(0,Y_i,t\right)-\Delta\phi_y\left(0,Y_{N+i},t\right)\right]
\end{eqnarray}
and compute the output signal $H_N\left(t\right)$ of the detector 
using Eq.~\eqref{average}.
It contains the GW signal $h(t)$, as well as the detection noise $\epsilon(t)$, and the  NN $a(X,t)$.
To derive the detector strain sensitivity curve, e.g. the minimum detectable GW power spectral density (PSD) $S_h(\omega)$ \cite{CreigthonBook,Moore2015}, we compute  the PSD  of the detector output, $S_{H_N}(\omega)$, using Eqs.\eqref{eq:gradient}, ~\eqref{average} and  \eqref{differential} : 
\begin{eqnarray}
\label{eq:S_Hn}
\nonumber S_{H_N}(\omega) & = & (2nk L)^2   \omega^4 S_h(\omega) |\hat{s}_\alpha(\omega)|^2  \\
& + &  (2nk)^2 S_a(\omega) |\hat{s}_\alpha(\omega)|^2  +  \frac{4 S_\epsilon(\omega)}{N}.
\end{eqnarray}
Here $\hat{s}_\alpha(\omega)=4\sin^2\left(\omega T/2\right)/\omega^2$ is the Fourier transform of the AI sensitivity function to acceleration $s_\alpha(t)$, and $S_\epsilon(\omega)$ is the PSD of the detection noise. The reduction by the factor $N$ reflects the uncorrelated detection noise in the different AIs.
The ratio between the first term (the GW contribution) and the last two terms (the noise PSD) of Eq.\eqref{eq:S_Hn} defines the SNR of our detection. 
If we consider a minimum sensitivity with a SNR of 1, we obtain the strain sensitivity function
\begin{equation}
S_h\left(\omega\right) = \frac{S_a(\omega)}{\omega^4 L^2}+\frac{4S_\epsilon(\omega)}{16NL^2\left(2nk\right)^2\sin^4\left(\omega T/2\right)}.
\label{eq:sensitivity_curve}
\end{equation}
The NN PSD $S_a(\omega)$ contains two contributions: one given by the gravity acceleration correlations between AIs at two positions $\{X_i,X_j\}$ in the same arm, and one given by the correlations between AIs at two positions $\{X_i,Y_j\}$ in orthogonal arms. 
The calculation of these contributions is detailed below.

\begin{figure}[!h]
  \centering
 \includegraphics[width=\linewidth]{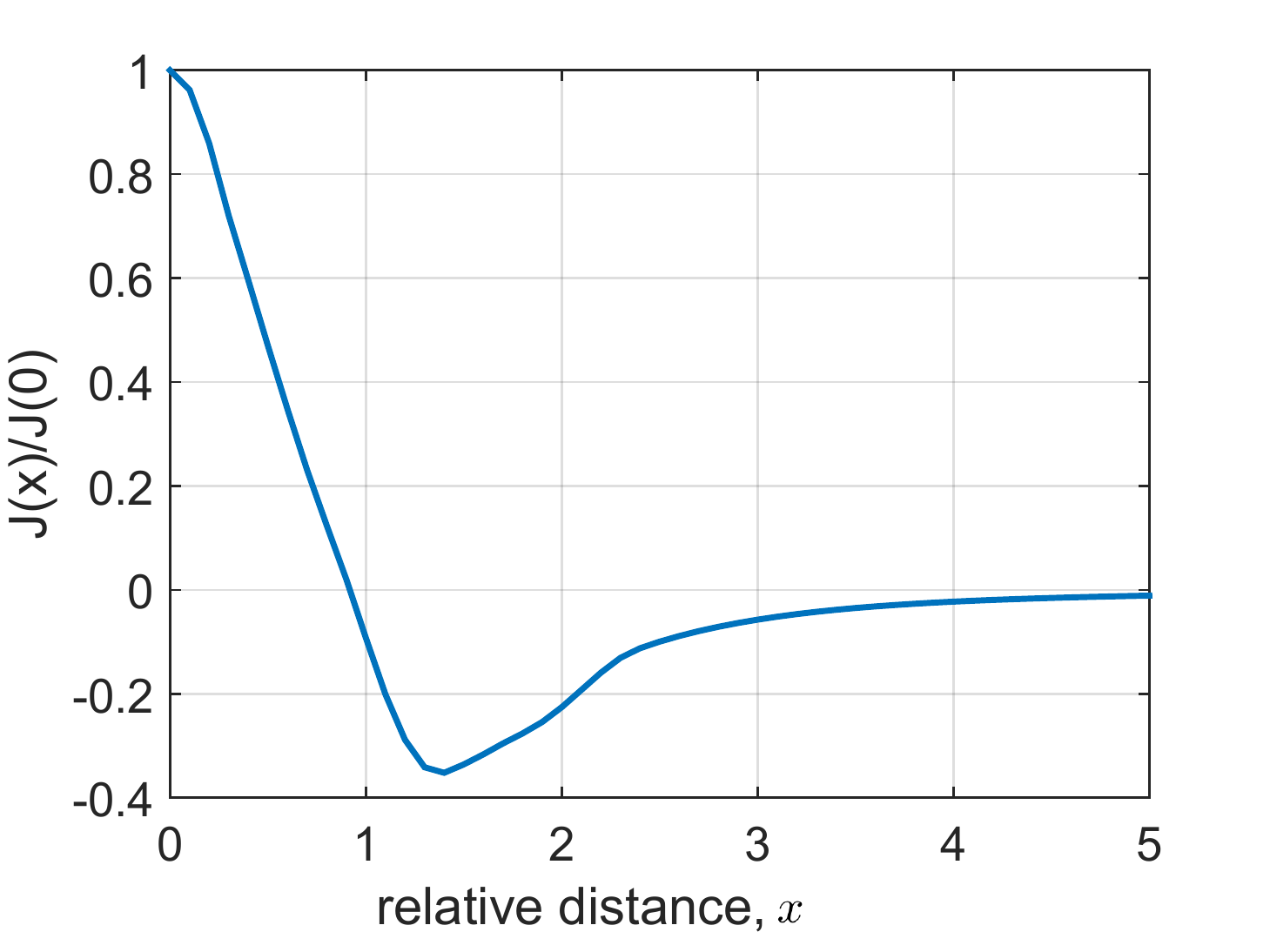}
  \caption{Spatial behavior of the normalized NN correlations between two distant points separated by the relative distance $x=|X_j-X_i|/\mathcal{L}_\rho(\omega)$, where $\mathcal{L}_\rho(\omega)$ is the  NN correlation length. 
  The anti-correlation is a consequence of mass conservation between adjacent cells of fluctuating density.
  }  
 \label{fig:J_aa}
  \end{figure}

Before looking into the details of the AI array rejection method, we review the sources of NN, which
are related to the modification of the mass distribution around the detector. We focus on the two main sources previously identified for ground based detectors:  
\textit{(i)} seismic noise related to elastic waves propagating within the ground~\cite{Saulson1984,Beccaria1998,Hughes1998} (seismically induced Newtonian Noise - SNN); \textit{(ii)} air mass fluctuations in the near atmosphere~\cite{Saulson1984,Creighton2008}. We base our calculation on the Saulson model~\cite{Saulson1984}: for each frequency $f=\omega/2\pi$, the ground is subdivided into  cells of fluctuating density  whose size corresponds to the half wavelength $\mathcal{L}_\rho(\omega)=v_u/2f$ of a propagating compression wave of velocity $v_u$. 
More specifically, we use an upgrade of the Saulson model that guarantees the mass conservation by assuming an anti-correlation between  adjacent cells \cite{Beccaria1998}. 
We plot in Fig.\ref{fig:J_aa} the spatial behavior of the  gravity acceleration correlation between two distant points.
Mass conservation yields a negative minimum of the correlation function for a characteristic length, which as been reported for the seismic noise in  Ref.~\cite{Mykkeltveit1983}.
The main other sources of low frequency NN  are those related to air pressure fluctuations caused by wind induced air turbulence \cite{Saulson1984} (Infrasound Newtonian Noise, INN), and  to the effect of turbulence induced frozen cells of random temperature dragged by the wind \cite{Creighton2008}. 
For a detector at depth $H$, the latter effect has a cut-off frequency $f_c = v_{\text{wind}}/\left(4\pi H\right)$ \cite{Creighton2008} which is out of the  detector band for $H > 100 \m$ ($v_{\text{wind}}\simeq 10-20\,\textrm{m/s}$ is the wind velocity).

We now give some details on the calculation of the NN contribution $S_a(\omega)$ appearing in Eq.~\eqref{eq:sensitivity_curve}, which we express as:
\begin{equation}
S_a\left(\omega\right)=\frac{1}{N^2} \sum^{2N}_{i,j=1}\mathcal{C}_{\parallel}\left(X_i,X_j,\omega\right)+\frac{1}{N^2} \sum^{2N}_{i,j=1}\mathcal{C}_{\perp}\left(X_i,Y_j,\omega\right),
\label{eq:S_NN}
\end{equation}
with the single arm component
\begin{equation}
\sum^{2N}_{i,j}\mathcal{C}_{\parallel}\left(X_i,X_j\right) \equiv 4 \sum^{N}_{i,j}C_{xx}(X_i,X_j) - 4 \sum^{N}_{i,j}C_{xx}(X_i,X_{j+N})
\label{eq:C_par}
\end{equation}
and the crossed arms component
\begin{eqnarray}
\label{eq:C_perp}
&&\sum^{2N}_{i,j}\mathcal{C}_{\perp}\left(X_i,X_j\right) \equiv -2 \sum^{N}_{i,j}C_{xy}(X_i,Y_j) \\
\nonumber
&&- 2 \sum^{N}_{i,j}C_{xy}(X_{i+N},Y_{j+N}) 
+ 4 \sum^{N}_{i,j}C_{xy}(X_{i+N},Y_{j}).
\end{eqnarray} 
In Eqs.~\eqref{eq:C_par} and \eqref{eq:C_perp},  $C_{xy}$ is the Fourier transform of the gravity acceleration correlation function between two AIs in arms $(x,y)$, and we hid the $\omega$ dependency for clarity.
We assumed isotropy of the NN and that the detector is surrounded by a homogeneous medium  for both seismic and infrasound-air density fluctuations. We also consider  the effects of the SNN and INN  as independent, so that the incoherent sum of the two contributions provides  an upper bound of our detector sensitivity.
With this model, the correlation $C_{xx}$ between two points in the same arm is given by 
\beq
C^{\NNtype}_{xx}\left(X_i,X_j,\omega\right) \simeq G^2 \mathcal{L}_{\rho}^{\NNtype}(\omega)^2 \Delta\rho_{\NNtype}^2(\omega)\times J\left(x_{ij}^{\NNtype}(\omega)\right), 
\label{eq:accel_correlations}
\eeq
with $x_{ij}^{\NNtype}(\omega)=\frac{\left|X_i-X_j\right|}{\mathcal{L}_{\rho}^{\NNtype}(\omega)}$.
Here $G$ is the gravitational constant, $\NNtype$ are indices denoting the seismic and infrasound NN  contribution and $\mathcal{L}_\rho^{\NNtype}(\omega) = \pi v_{u,a}/\omega$ is the corresponding  correlation length, with $v_u$ and $v_a$ being respectively the speed of seismic waves in the underground and the speed of sound in the air.
The function $J\left(x\right)$ is a 3D integral which represents the  spatial behaviour of NN correlations between two distant points $X_i$ and $X_j$. It  is represented in Fig.~\ref{fig:J_aa} against the relative distance $x$.
A similar expression as Eq.~\eqref{eq:accel_correlations} holds for $C_{xy}$, the correlation  between two points $\{X_i,Y_j\}$ in orthogonal arms.

Following Refs.~\cite{Saulson1984,Harms2013}, the density fluctuations for SNN and INN are respectively given by
$\Delta\rho_u^2(\omega) = \frac{\rho_u^2\Delta a_s^2(\omega)}{\pi\omega^2 v_u^2}$ and
$\Delta\rho_a^2(\omega) = \frac{\rho_a^2}{\gamma^2 p_a^2} \Delta p^2(\omega)$. Here $\rho_u = 2300\,\textrm{kg/m}^3$ is the mean underground density, $\Delta a_s(\omega)$  the seismic acceleration noise, $\rho_a=1.3\,\textrm{kg/m}^3$  the mean air density, $1/\gamma^2\simeq 1/2$  the air coefficient of adiabatic compression, $p_a$  the air pressure and $\Delta p^2(\omega)$ its PSD.
We consider seismic waves with typical speed for P waves $v_u = 2 \ \text{km/s}$ corresponding for example to porous rocks \cite{MavkoStanford}, yielding $\mathcal{L}_\rho= 1 \ \text{km}$ at 1 Hz.
The air pressure fluctuation spectrum used for the INN is $\Delta p^2(\omega)=0.3 \times 10^{-5}/(f/1\Hz)^2\  \text{Pa}^2/\text{Hz}$   (as used by Saulson \cite{Saulson1984}). The seismic noise for the SNN is $1\times 10^{-17} \ \text{m}^{2}\text{s}^{-4}/\text{Hz}$ at $1 \Hz$ as often reported in  underground sites, see e.g. \cite{Farah2014}.

The gradiometer separation $\delta$  determines  the NN rejection efficiency.
For instance, if 
$\delta$ is much larger than the NN correlation length 
$\mathcal{L}_{\rho}(\omega)$ for all $\omega$, 
then the successive measurement points are uncorrelated and Eq.~\eqref{eq:S_NN} reduces to terms $i=j$, yielding:
\begin{equation}
S_a^{\NNtype}\left(\delta_\infty,\omega\right) \simeq \frac{4}{N}G^2 \mathcal{L}_{\rho}^{\NNtype}(\omega)^2 \Delta\rho_{\NNtype}^2(\omega)\times J\big(0\big).
\label{eq:infinite_delta}
\end{equation}
This situation, which corresponds to the standard Monte-Carlo method (see Eq.~\eqref{eq:MCnoisered}), already determines a significant NN rejection of $\sqrt{N}$ (in noise amplitude). 
Choosing an optimal value for $\delta$,  it is then possible to benefit from the anti-correlation in the NN (corresponding to negative values in Fig.~\ref{fig:J_aa}). In this case, the Monte-Carlo variance reduction \cite{Caflisch1998} increases the NN rejection of Eq.~\eqref{eq:infinite_delta}. 
The choice of the AI array sampling pattern (i.e. $\delta$, $\delta_0$ and $\delta_N$) sets the correlation between the measurement points  and thus the amount of additional NN rejection compared to $\sqrt{N}$.
The INN and SNN rejection prefactors depend on the shape of $J(x)$, i.e. on the characteristics of the site \cite{Braun2008}. 

\begin{figure}[!h]
\centering
 \includegraphics[width=\linewidth]{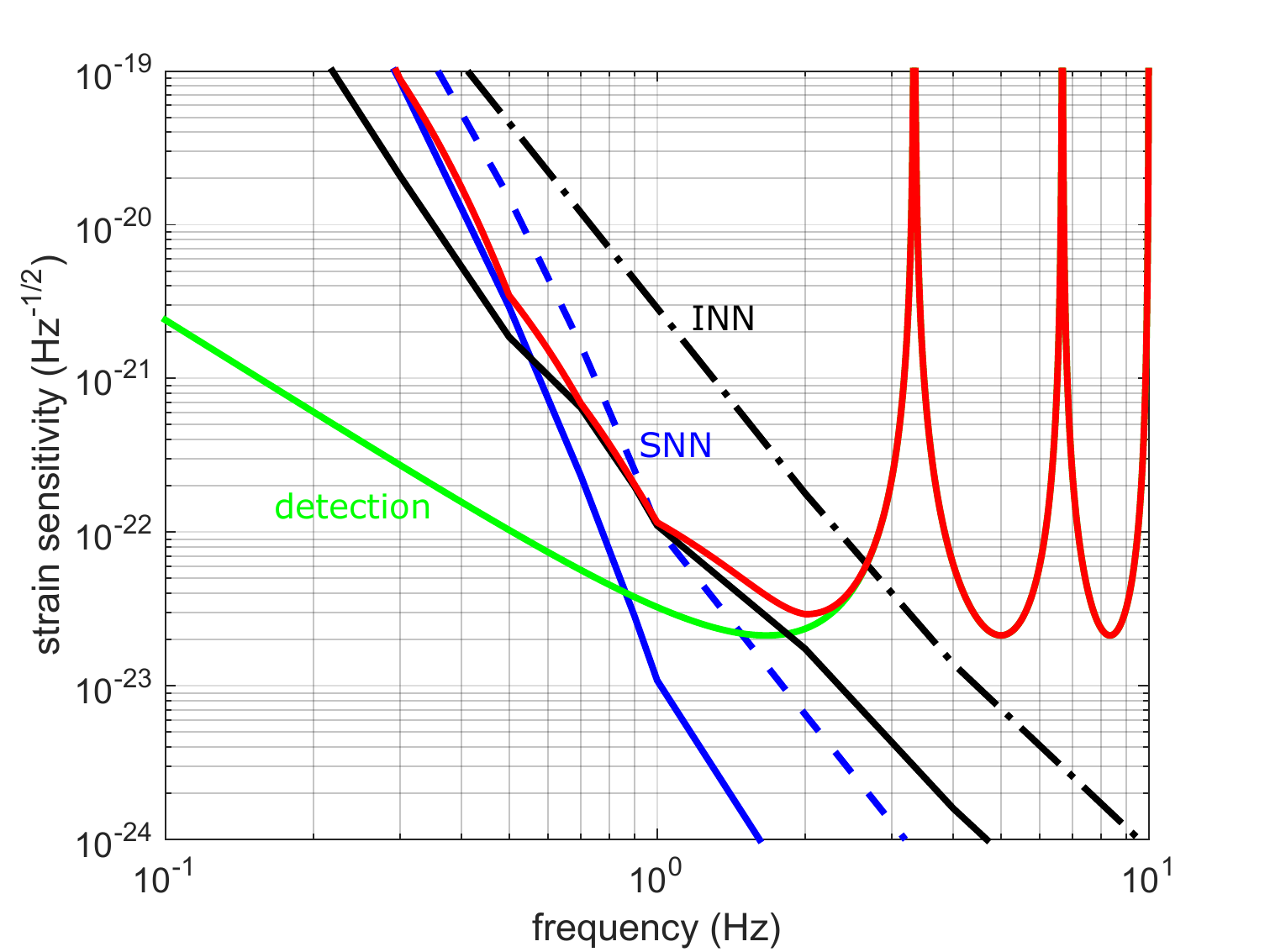}
 \caption{(Color)  Strain sensitivity curve for an AI array with $N = 80$, $\delta = 200 \m$, $\delta_0=\delta_N = 500 \m$, $L=16.3 \km$ and $L_a = 32.6 \km$.
The AI phase noise is $-140 \textrm{dB rad}^2/\textrm{Hz}$ with the interrogation time $T = 0.3 \s$, and $n = 1000$ LMT beam splitters.   Green: detection noise; dotted-dashed black (dashed blue): INN (SNN) for two test masses separated by the baseline $L$; solid black line (blue): residual INN (SNN) after NN rejection with the AI array. Red: overall sensitivity curve. 
}
 \label{fig:sensitivity_curve}
  \end{figure}

We illustrate our discussion with a configuration of $N=80$ gradiometers of baseline $L=16.3  \ \text{km}$, separated by the distance $\delta = 200 \m$.
We plot the expected strain sensitivity function in Fig.~\ref{fig:sensitivity_curve}, using Eqs.~\eqref{eq:sensitivity_curve}--\eqref{eq:C_perp}. 
We use a detection noise PSD $S_\epsilon=-140 \ \textrm{dB rad}^2/\textrm{Hz}$ which corresponds, for example, to $N_{at}= 10^{12}$  atoms per second and a $20\,\text{dB}$ reduction (in variance) in the detection phase noise by using entangled atomic states. 
We assumed LMT beam splitters with $n=1000$. 
Similar parameters have been considered in other AI proposals  (see, e.g. \cite{Dimopoulos2008PRD}).
The total AI interrogation time is chosen to $2T = 0.6 \ \text{s}$, which is compatible with the high sampling frequencies and the absence of dead times required for GW detection by using  joint interrogation sequences \cite{Meunier2014}.

The NN reduction offered by the  AI array is maximal around  $1 \Hz$ where it exceeds 30 for the INN and 10 for the SNN, yielding a shot noise limited strain sensitivity level of $3\times 10^{-23} \sqrtHz$ at $2 \Hz$.
At low frequency ($\lesssim 0.3 \Hz$), the SNN correlation length becomes much greater than  $\delta$ which results in a high correlation between the different gradiometer measurements, thereby preventing the NN rejection. 
At high frequencies ($> 2\Hz$), the detector is limited by detection noise.

\section{Discussion.}
\label{par:discussion}

\begin{figure}[!h]
  \centering
 \includegraphics[width=\linewidth]{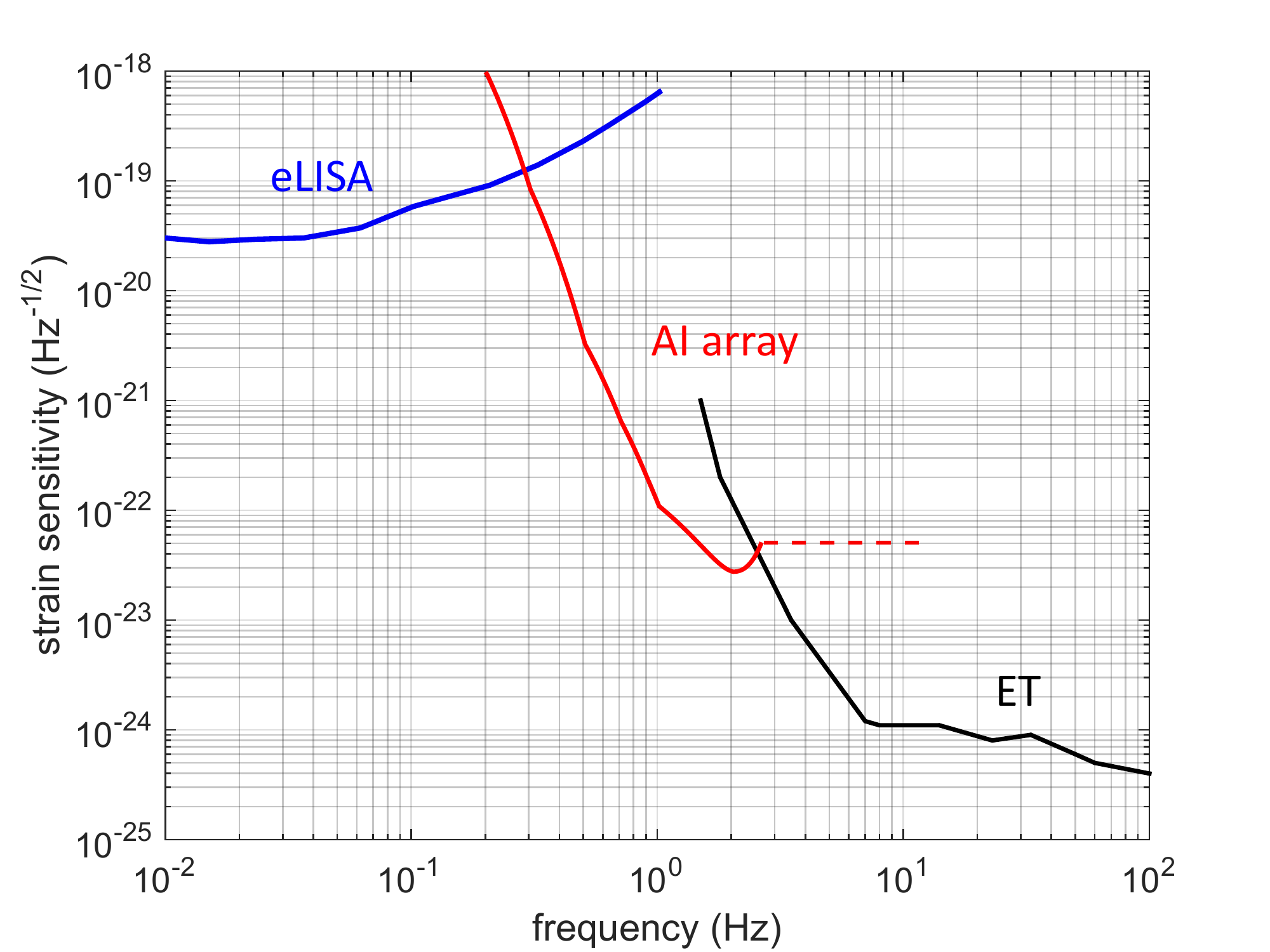}
  \caption{The strain sensitivity of the proposed AI array covers  the frequency region  $\sim 0.3-3 \Hz$, where future ground-based (Einstein Telescope - ET) and space-based (eLISA) detectors are blind. 
The dashed line represents an envelope  of the proposed AI array sensitivity function at frequencies above $3 \Hz$ and corresponds to an average detector response for different interrogation times $T$.
  }
 \label{fig:comparison_AI_others}
  \end{figure}

As shown in Fig.~\ref{fig:comparison_AI_others}, such performances  would allow  observations in the frequency band $\sim 0.3-3 \Hz$. This frequency band is  covered neither  by existing detectors nor by  next generation detectors such as the Einstein Telescope \cite{Punturo2010} or ESA's L3 gravity observation mission eLISA \cite{LISA_sensitiv_curve}, despite the presence of several  astrophysical sources \cite{Adhikari2014}. 


To conclude, we show that an array of AIs in an appropriate configuration can allow ground based GW detection in the  $\sim 0.3-3 \Hz$ decade by overcoming the current limitation imposed by NN. 
The main idea consists in using a distribution of long baseline AI gradiometers  to average the NN to zero.
We  show that a further NN reduction can be achieved   by exploiting the  NN correlation properties to configure the AI array.
While the present concept can be tested on existing apparatuses, our method will take full advantage on the recent and future development in atom interferometry. More advanced schemes might also lead to sensitivity improvements.
For example, the measurement of higher order spatial derivatives of the gravity field \cite{Rosi2014curvature}, or the implementation of more complex spatial distributions of AIs  could achieve higher NN rejections, depending on the site-dependent  NN correlations.
Detectors based on AI arrays  could then help filling the blind frequency band between ground-based and space-based detectors.

\section{Acknowledgements}
\label{par:acknowledgements}
We acknowledge P. Delva, P. Wolf, B. Chauvineau, J.-Y. Vinet and T. Regimbau for discussions, and financial support from the MIGA Equipex funded by the French National Research Agency (ANR-11-EQPX-0028).

\section{Appendix}
\label{appendix}

\subsection{Requirement on the seismic isolation of the beam splitting optics}
\label{ap:seismic_iso}
In the proposed configuration consisting of 2 orthogonal arms  (Fig.\ref{fig:schematic_detector}(b)), the beam splitting optical system  that distributes the laser to the 2 arms introduces an asymmetry. 
Position noise (e.g. seismic noise) of the splitting optics results in laser frequency noise which will affect one arm and not the other: the phase $\varphi_L$ of the  laser beam propagating in the $y$ direction  picks up the position noise $\delta y$ of the splitting optics, which results in a frequency noise contribution $\Delta\nu = \frac{1}{2\pi}\frac{d\varphi_L}{dt}=k f \delta y$ (with Fourier frequency $f$). 
According to Eq.~\eqref{phase_AI}, such frequency noise  yields a contribution to the relative phase signal of the AI gradiometers in the $y$ arm equal to $2nkL\times \Delta\nu(t)/\nu =2nkL\times 2\pi f\delta y /c$, to be compared with the GW signal $2nk\times Lh$.
Considering a minimum sensitivity with a SNR of 1  yields the requirement on the position noise $\delta y\stxt{min}$ of the splitting optics given by $\delta y\stxt{min}=hc/2\pi f$. 
To reach a detector peak sensitivity of $3\times 10^{-23}\sqrtHz$ at $f=2\Hz$, the seismic noise must be below $\delta y\stxt{min} (2\Hz)\approx 7\times 10^{-16} \ \text{m}\sqrtHz$. At $f=0.3\Hz$, the AI array can feature a sensitivity of $1\times 10^{-19}\sqrtHz$ if the seismic noise is mitigated below $\delta y\stxt{min} (0.3\Hz)\approx 2\times 10^{-11} \ \text{m}\sqrtHz$.
Such seismic noise levels can be obtained with a dedicated low frequency suspension system (see, e.g. \cite{Liu1997}).
Finally, the contribution resulting from NN induced position fluctuations of the splitting optics is negligible at the targeted sensitivity level.

\subsection{Newtonian Noise rejection efficiency}
\begin{figure}[!h]
\centering
 \includegraphics[width=\linewidth]{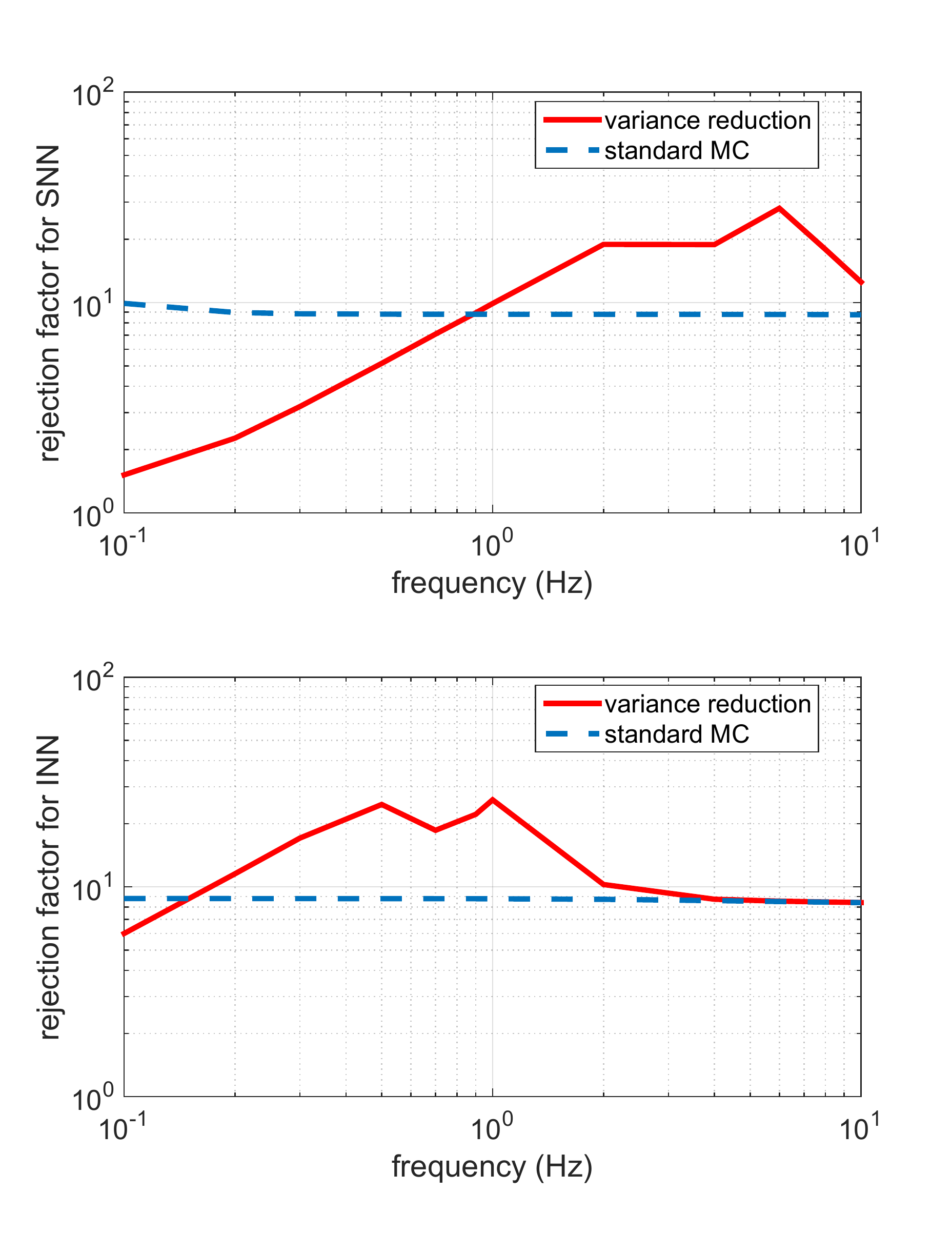}
 \caption{Noise rejection factor of the SNN (top) and the INN (bottom) for the  implementation of the AI array described in the main text. 
}
 \label{fig:rejection_efficiency}
  \end{figure}

Fig.~\ref{fig:rejection_efficiency} illustrates the NN rejection efficiency of the AI array. The dashed line shows the rejection in the case of a standard Monte Carlo average illustrating the $\sqrt{80}$ rejection factor. The plain line shows the rejection using the Monte Carlo variance reduction method exploiting the spatial behavior of the gravity acceleration correlation function. 
The maximum rejection is obtained when the NN correlation length $\mathcal{L}_\rho^{\NNtype} =  v_{u,a}/2f$ approches the distance corresponding to the anti-correlation of the gravity acceleration correlation function, which, from Fig.~\ref{fig:J_aa},  is obtained for $x\stxt{ac}\approx 1.3$. This condition on the length translates in the frequency where the maximum rejection is observed, given by $f= v_{u,a} x\stxt{ac}/2\delta$ and equals $ 1.1 \Hz$ for the INN and $ 6.5 \Hz$ for the SNN.

\bibliographystyle{apsrev4-1}
\bibliography{beating_NN_biblio}


\end{document}